\newcommand\strt[1]{\rule[-#1pt]{0pt}{#1pt}}
\def\bnab{{\boldsymbol \nabla}}
\documentclass[proof]{WileyASNA-v1}

\articletype{Article Type}%

\received{0/00/2019}
\revised{0/00/2019}
\accepted{0/00/2019}

\begin{document}

\title{Local Metric with Parameterized Evolution}

\author{Martin Land}


\address{\orgdiv{Department of Computer Science}, \orgname{Hadassah College}, \orgaddress{\country{Jerusalem}}}

\corres{\email{martin@hac.ac.il}}


\abstract{We present a canonical Hamiltonian formulation of GR in which $\tau$,
the parameter of system evolution, is external to spacetime, playing a role
similar to what we call time in nonrelativistic mechanics. 
This approach, known as Stueckelberg-Horwitz-Piron (SHP) theory, 
inherits the full computational power of classical analytical mechanics
while maintaining manifest covariance throughout and eliminating possible
conflict with general diffeomorphism invariance.  
In particular, SHP 
simplifies the initial value problem
with potential applications in highly dynamical interactions, such as black hole
collisions.  
By allowing the energy-momentum tensor and metric to depend explicitly on
$\tau$, we may describe particle motion in geodesic form with respect to a
dynamically evolving background metric. 
As a toy model, we consider a $\tau$-dependent mass $M(\tau)$, first as a
perturbation in the Newtonian approximation and then for a Schwarzschild-like
metric. 
As expected, the extended Einstein equations imply a non-zero energy-momentum
tensor, proportional to $dM / d\tau$,
representing a flow of mass and energy into spacetime that corresponds to
the changing source mass.  
In $\tau$-equilibrium, this system becomes a generalized Schwarzschild solution
for which the extended Ricci tensor and energy-momentum tensor vanish.}  


\maketitle

\section{Introduction}\label{sec1}

In general relativity, matter tells spacetime how to curve and spacetime
curvature tells matter how to move.  
More precisely, a mass distribution given by 
$T^{\alpha\beta} \left( x \right) $ induces a local metric
$g_{\alpha\beta}\left( x \right)$ through the Einstein equations, and the metric
enters the equations of motion for particles.
One may, in principle, construct the $T^{\alpha\beta} \left( x \right)$
associated with 
particles $x^\mu_i \left( \tau \right)$,
find the induced metric and
obtain equations of motion for matter in the resulting gravitational field.
But calculations of this type may be intractable --- even for 
linearized field equations. 
As Fock 
observed~\cite{Fock}, it may be impractical to replace $\tau$ by $t$ as parameter
of particle motion, in order to obtain $T^{\alpha\beta}$ as a function of
$x^\mu$ alone.  Later, Stueckelberg~\cite{Stueckelberg-1,Stueckelberg-2} argued
that to fully characterize all possible spacetime trajectories, the 8D phase space
$x^\mu \left( \tau \right) , \dot x^\mu \left( \tau \right)$ must be
unconstrained, and
$\tau$ cannot be identified with the proper
time of the motion, which may not be well-defined.  Horwitz and Piron~\cite{HP}
developed this idea into a manifestly covariant canonical mechanics for the
many-body problem~\cite{rel-qm,MB}, in which gauge fields may depend explicitly on
$\tau$ so that the Stueckelberg-Horwitz-Piron~(SHP) theory remains well-posed~\cite{saad}.

The classical and quantum mechanics of particles in a spacetime with local
metric $g_{\mu\nu}(x)$ has been studied extensively by Horwitz~\cite{SHPGR-1} and will not be
discussed at length here.  Our goal is to find a consistent prescription for
extending general relativity to accommodate an energy-momentum
tensor $T^{\alpha\beta} (x,\tau) $ and a metric $g_{\alpha\beta}(x,\tau)$ 
satisfying $\tau$-dependent Einstein equations.  
As a guide we recall the construction of the SHP particle
action as $S_{\text{Maxwell}} \longrightarrow S_{\text{SHP}}$, where
\begin{equation}
S_{\text{Maxwell}} =\int d\tau ~\frac{1}{2}M\dot{x}^{\mu }\dot{x}_{\mu }+\frac{e}{c}\dot{x}
^{\mu }A_{\mu }\left( x^{\lambda }\right)
\end{equation}
for $\lambda , \mu , \nu =0,1,2,3$, and 
\begin{eqnarray}
S_{\text{SHP}}
& = &  \int d\tau ~\frac{1}{2}M\dot{x}^{\mu }\dot{x}_{\mu }+\frac{e
}{c}\dot{x}^{\mu }a_{\mu }\left( x^{\lambda },\tau \right) 
+\frac{e}{c} c_{5}a_{5}\left( x^{\lambda },\tau \right)
\label{SM} \notag \\
& = &  \int d\tau ~\frac{1}{2}M\dot{x}^{\mu }\dot{x}_{\mu }
+\frac{e }{c}\dot{x}^{\beta }a_{\beta }\left( x^{\lambda },\tau \right)
\label{SSHP}
\end{eqnarray}
where $\alpha,\beta,\gamma = 0,1,2,3,5$, and in analogy to $x^0 = c t$, we
write $x^5 = c_5 \tau$.  Variation with respect to $x^\mu$ leads to the Lorentz
force~\cite{lorentz} 
\begin{eqnarray}
M\ddot{x}_{\mu } & = &  \frac{e}{c}\left( \dot{x}^{\nu }f_{\mu \nu }+c_{5}f_{\mu
5}\right) = \frac{e}{c}\dot{x}^{\beta }f_{\mu \beta } 
\label{L-1}
\\
\frac{d}{d\tau }\left( - \frac{1}{2}M\dot{x}^{\mu }\dot{x}_{\mu }\right)
& = &   c_{5}\frac{e}{c}\dot{x}^{\beta }f_{5 \beta }
\label{L-2}
\end{eqnarray}
where (\ref{L-2}) permits the exchange of mass between particles and fields.
The $\beta = 5$ index is understood to be a formal convenience, indicating
O(3,1) scalar quantities, and not elements of 5D tensors.  Still, terms of the
type $\dot{x}^{\beta }a_{\beta }\left( x,\tau \right)$ suggest that (\ref{L-1})
and (\ref{L-2}) exhibit $5D \longrightarrow 4D$ symmetry breaking, in which we
first specify $\dot x^5 = c_5$ as a scalar constant, restricting the dynamical
degrees of freedom to the 4D manifold, {\em before} applying a dynamical
principle.  

\section{Canonical mechanics}
\label{canon}

\subsection{An approach to a $\tau$-dependent metric}
\label{approach}

Standard general relativity considers the invariant interval
\begin{equation}
\delta x^2 = g_{\mu\nu} \delta x^\mu \delta x^\nu = \left( x_2 - x_1 \right)^2 
\label{interval}
\end{equation}
between two neighboring spacetime points, viewed as an instantaneous displacement. 
This invariance is a geometrical statement expressing a wide freedom in
assigning coordinates to the manifold. 
To transform geometry into dynamics, a spacetime
path maps an arbitrary parameter
$\zeta$ to a continuous sequence of events $x^\mu(\zeta)$ such that 
any two points on the path have timelike separation, and so the 
proper time can be taken as the parameter.  Consistent with the notion of a 4D
block universe, the path consists of instantaneous displacements, observed as
``motion'' through changes in $x^0(\zeta)$ with $\zeta$. 
Parameterizing by proper time $s$,  
the line element
\begin{equation}
\delta x^2 = g_{\mu\nu} \delta x^\mu \delta x^\nu
= g_{\mu\nu} \frac{dx^\mu}{ds}\frac{dx^\nu}{ds}  \delta s^2
= g_{\mu\nu} \dot x^\mu \dot x^\nu  \delta s^2 
\label{interval-2}
\end{equation}
suggests a dynamical description of the path by the action
\begin{equation}
S = \int dx  = \int ds \ \sqrt{-g_{\mu\nu} \dot x^\mu \dot x^\nu }
\label{sqrt}
\end{equation}
and leading to geodesic equations of motion as an expression of the equivalence
principle.  The geodesic equations can also be derived from the action
\begin{equation}
S = \int ds \ \frac{1}{2} \ g_{\mu\nu} \dot x^\mu \dot x^\nu 
\end{equation}
which removes the constraint $\dot x^2 = -c^2$ associated with (\ref{sqrt}).

In SHP a physical event $x^\mu (\tau)$ theory occurs {\em at} time $\tau$ and
chronologically precedes events occurring at subsequent times.  
The physical picture 
describes the evolution of one 4D block universe defined at time
$\tau$ to another, infinitesimally close, at $\tau + d\tau$.
Evolution slows to zero in equilibrium.
The $\tau$-dependence of $a^\alpha (x,\tau)$ reminds us
that while geometric relations on spacetime are defined within a given block universe, the
dynamics are defined in the transition from one 4D block manifold to another.  
We therefore consider the interval
\begin{equation}
dx^\mu =  \bar x^\mu (\tau + \delta \tau) - x^\mu (\tau)
\end{equation}
between an event $x^\mu$ occurring at time $\tau$ and an event $\bar x ^\mu$
occurring at a different spacetime location at a subsequent time $\tau + \delta
\tau$.  
With $\delta x^5 = c_5 \delta \tau$ we may write
\begin{equation}
dx^2 
=  g_{\alpha\beta} \left( x,\tau\right)  \delta x^\alpha \delta
x^\beta 
\end{equation}
which contains both the geometrical
distance $\delta x^\mu$ between two neighboring points in one manifold, and the
dynamical distance $ \delta x^5$ between events occurring at two sequential
times.  

This line element suggests the free particle Lagrangian 
\begin{equation}
L = \frac{1}{2} M g_{\alpha\beta} \left( x,\tau\right)  \dot x^\alpha \dot x^\beta
\label{Lag-1}
\end{equation}
containing both geometry and dynamics.  In a formal and restricted sense, this
system is defined on 5D coordinates $\left( x^\mu, x^5\right)$,
as was done for the fields in (\ref{SM}) to (\ref{L-2}). 
Nevertheless, we maintain the distinction between geometry and evolution. 
The parameter $x^5$ does not become a dynamical degree
of freedom, and no five dimensional symmetries are ascribed to this 5D
structure.
While the terms $c_5 g_{\mu 5} \left( x,\tau\right)  \dot x^\mu$ and $c_5^2 g_{55}
\left( x,\tau\right) $ appear as 
interactions, we view them
as features of an extended equivalence principle, in which
$g_{\mu\nu}$ expresses geodesic motion within the curved geometry of a 4D block
universe at any fixed $\tau$, while $g_{\alpha 5}$
express geodesic motion along the evolving structure of one 4D block universe
to another.\footnote{Horwitz and Gershon \cite{gershon} have shown that a covariant
action with scalar potential is equivalent to a free action with local conformal
metric.  In their work, metric component $g_{5\alpha}$ and related components of
the connection similarly appear.}

\subsection{General 5D spacetime}
If (\ref{Lag-1}) were a 5D free particle Lagrangian
for $ x^\alpha (\xi) $ then the
Euler-Lagrange equations
would provide geodesic equations
\begin{equation}
0 =\frac{D\dot{x}^{\gamma}}{D\xi}=\ddot{x}^{\gamma}+\Gamma _{\alpha\beta}^{\gamma}\dot{x}^{\alpha}\dot{x} ^{\beta}
\label{motion-1}
\end{equation}
where $D/D\xi$ is the absolute derivative (in the notation of Weinberg \cite{Weinberg}) and
\begin{equation}
\Gamma _{\alpha \beta}^{\gamma }=g^{\gamma \delta}\Gamma _{\delta \alpha
\beta}=\frac{1}{2}g^{\gamma \delta}\left( \partial _{\alpha }g_{\delta
\beta}+\partial _{\beta}g_{\delta \alpha}-\partial _{\delta
}g_{\beta\alpha}\right) 
\label{connection}
\end{equation}
is the standard Christoffel symbol in 5D. 

Writing the canonical momentum 
\begin{equation}
p_{\alpha} = \frac{\partial L}{\partial \dot{x}^{\alpha}}=Mg_{\alpha\beta}\dot{x}^{\beta} \qquad
\longrightarrow \qquad 
\dot{x}^{\alpha}  = \frac{1}{M}g^{\alpha\beta}p_{\beta}
\end{equation}
the scalar Hamiltonian, representing particle mass,
\begin{equation}
K=\dot{x}^{\alpha}p_{\alpha}-L = \frac{1}{2M}g^{\alpha\beta}p_{\alpha}p_{\beta} = L
\end{equation}
is conserved, as seen directly through
\begin{equation}
\frac{d}{d\tau}\left(
\frac{1}{2}Mg_{\alpha\beta}\dot{x}^{\alpha}\dot{x}^{\beta}\right) 
=Mg_{\alpha\beta}\dot{x}^{\alpha}\frac{D\dot{x}^{\beta}}{D\tau}=0
\end{equation}
where we used metric compatibility $Dg_{\alpha\beta} / D\xi = 0$.
This result may also be seen from the canonical equations of motion
\begin{equation}
\dot{x}^{\alpha}=\frac{dx^{\alpha}}{d\xi }=\frac{\partial K}{\partial p_{\alpha}}\qquad
\qquad \dot{p}_{\alpha}=\frac{dp_{\alpha}}{d\xi }=-\frac{\partial K}{\partial x^{\alpha}}
\end{equation}
and the Poisson bracket
\begin{equation}
\left\{ F,G\right\} =\frac{\partial F}{\partial x^{\alpha}}\frac{\partial G}{
\partial p_{\alpha}}-\frac{\partial F}{\partial p_{\alpha}}\frac{\partial G}{\partial
x^{\alpha}} 
\end{equation}
so that
\begin{equation}
\frac{dK}{d\xi}=\left\{ K,K\right\} +\frac{\partial K}{\partial \xi}
 =  \frac{1 }{2M}  p_{\alpha}p_{\beta} \ \frac{\partial g^{\alpha\beta}}{\partial \xi} =0 
\end{equation}
using the $\xi$-independence of the metric.

\subsection{Breaking the 5D symmetry to 4D+1}

We break the 5D symmetry by putting
\hbox{$x^5(\tau) = c_5 \tau$} in (\ref{Lag-1}), which becomes  
\begin{equation}
L 
= \frac{1}{2}Mg_{\mu\nu}\dot{x}^{\mu}\dot{x}^{\nu}
+ \frac{1}{2}Mc_5 g_{5\mu}\dot{x}^{\mu}
+ \frac{1}{2}Mc^2_5 g_{55} \ .
\label{Lag-2}
\end{equation}
Comparing (\ref{motion-1}) with the equations of motion 
\begin{eqnarray}
0 
& = & \ddot{x}^{\mu}+\Gamma _{\lambda \sigma
}^{\mu }\dot x^\lambda \dot x^\sigma +2c_5\Gamma _{5\sigma
}^{\mu }\dot x^\sigma +c^2_5\Gamma _{55}^{\mu }
\label{motion-3a}
\\
0 & = &  \frac{D\dot{x}^{5}}{D\tau} = \frac{d\dot{x}^{5}}{d\tau}=\frac{dc_5}{d\tau}
\label{motion-3b}
\end{eqnarray}
the prescription for symmetry breaking connection is
\begin{equation}
\Gamma_{5\alpha}^{\mu} =
\frac{1}{2}g^{\mu \nu }\left( \partial _{5  }g_{\nu 
\alpha }+\partial _{\alpha }g_{\nu  5 }-\partial _{\nu 
}g_{\alpha 5 }\right)
\qquad \Gamma_{\alpha\beta}^5  \equiv 0 \ .
\label{prescription}
\end{equation}
We notice in particular the violation of the 5D symmetry
\begin{equation}
\Gamma _{5\mu\nu} =-\Gamma _{\mu 5\nu}+\partial _{\nu}g_{\mu 5}
\end{equation}
which no longer applies because the label 5 indicates a scalar quantity, and is
not a tensor index.

Because $x^5$ is not a dynamical quantity and has no conjugate momentum,
the Hamiltonian can be written 
\begin{equation}
K = \frac{1}{2}Mg_{\mu\nu}\dot{x}^{\mu}\dot{x}^{\nu} - \frac{1}{2}Mc^2_5 g_{55}
\label{Ham-2}
\end{equation}
which is not conserved for a $\tau$-dependent metric, as
seen from
\begin{equation}
\frac{d K}{d \tau }  
 = \frac{\partial K}{\partial \tau }  =  -\frac{1}{2}M\dot{x}^{\mu }\dot{x}^{\nu }\frac{\partial g_{\mu
\nu }}{\partial \tau }-\frac{1}{2}Mc_{5}^{2}\frac{\partial g_{55}}{
\partial \tau } .
\label{non-cons}
\end{equation}
Expression (\ref{non-cons}) can also be found by taking the total
\hbox{$\tau$-derivative} of (\ref{Ham-2}) and inserting the equations of motion
(\ref{motion-3a}). 

\subsection{Einstein equations}

We define $n(x,\tau)$ to be the number of events
per spacetime volume, so that the total number of events is
\begin{equation}
N(\tau) = \int d\Omega \ n(x,\tau) 
\end{equation}
where $d\Omega$ is a scalar spacetime volume element, and  
\begin{equation}
j^{\alpha }\left( x,\tau \right) =\rho(x,\tau) \dot{x}^{\alpha }(\tau) =M
n(x,\tau)\dot{x}^{\alpha }(\tau)
\end{equation}
is the 5-component event current.  The continuity equation in flat space 
\begin{equation}
\partial_\alpha j^\alpha = \partial_\mu j^\mu  + \partial_5 j^5 
= \partial_\mu j^\mu  + \frac{\partial \rho}{\partial \tau} =0
\label{continuity}
\end{equation}
leads to conservation of total event number through
\begin{equation}
\frac{d}{d\tau} \int d\Omega \ n(x,\tau) +
\frac{1}{M} \int d\Omega \ \partial_\mu j^{\mu }
= \frac{dN}{d\tau} =0  \ .  
\end{equation}
With a local metric (\ref{continuity}) is generalized (in the notation of Wald \cite{Wald}) to
\begin{equation}
{\boldsymbol \nabla}_{\alpha }j^{\alpha } = \bnab_\alpha X^\alpha =
\partial_\alpha j^\alpha + j^\gamma
\Gamma^\alpha_{\gamma\alpha} = 0
\end{equation}
and since $j^5$ is a scalar on physical grounds, 
prescription (\ref{prescription}) requires $\Gamma_{\alpha\beta}^{5} =0$, so
that the continuity equation becomes 
\begin{equation}
{\boldsymbol \nabla}_{\alpha }j^{\alpha } =
\frac{\partial \rho}{\partial \tau} + {\boldsymbol \nabla}_{\mu }j^{\mu } = 0 \ .
\end{equation}
Generalizing the 4D stress-energy-momentum tensor to 5D, we write the
mass-energy-momentum tensor \cite{antenna} as
\begin{equation}
T^{\alpha \beta }=Mn\dot{x}^{\alpha }\dot{x}^{\beta }=\rho \dot{x}^{\alpha }
\dot{x}^{\beta }
\end{equation}
where in addition to the standard 4D components $T^{\mu\nu}$, we have the current density
$T^{5\beta}=\dot{x}^{5}\dot{x}^{\beta}\rho =c_{5}j^\beta$.
Conservation of mass-energy-momentum $\bnab_\beta T^{\alpha \beta } = 0$
follows from the geodesic equations
$ \dot{x}^{\beta }\bnab_\beta \dot{x}^{\alpha }=D\dot{x}^\alpha  / D\tau =0$ and continuity 
${\boldsymbol \nabla}_{\alpha }j^{\alpha } =0$,
when the equations of motion (\ref{motion-3a}) and (\ref{motion-3b}) are
evaluated under the prescription (\ref{prescription}).   

The Einstein equations are similarly extended to 
\begin{equation}
G_{\alpha \beta} =
R_{\alpha \beta} - \frac{1}{2} Rg_{\alpha \beta} = \frac{8\pi G}{c^4} T_{\alpha \beta}
\label{Einstein}
\end{equation}
where the Ricci tensor $R_{\alpha \beta}$ and scalar $R$ are found by contracting
indices of the curvature tensor $R_{\gamma \alpha \beta}^\delta$
found from the Christoffel symbols.  Conservation of 
$T_{\alpha \beta}$ depends on
the prescription (\ref{prescription}), and so in order to insure 
\begin{equation}
\bnab_\beta G^{\alpha \beta} = 0
\end{equation}
we must similarly suppress $\Gamma_{\alpha\beta}^{5}$ when constructing the Ricci tensor.
Writing $R_{\alpha\beta}=R_{\alpha\beta\gamma}^{\gamma}=R_{\alpha\beta\lambda
}^{\lambda  }+R_{\alpha\beta 5 }^{ 5 }$,
we proceed by separating out terms containing the 5 index, leading to 
\begin{eqnarray}
R_{\mu \nu } & = &  \left( R_{\mu \nu }\right) ^{4D}
\label{ricci-1}
\\
R_{\mu  5 } & = &  \frac{1}{c_{ 5 }}\partial _{\tau }\Gamma _{\mu \lambda  }^{\lambda 
}-\partial _{\lambda  }\Gamma _{\mu  5 }^{\lambda  }+\Gamma _{\sigma   5 }^{\lambda 
}\Gamma _{\mu \lambda  }^{\sigma  }-\Gamma _{\sigma  \lambda  }^{\lambda  }\Gamma
_{\mu  5 }^{\sigma  } 
\label{ricci-2}
\\
R_{ 5  5 } & = &  \frac{1}{c_{ 5 }}\partial _{\tau }\Gamma _{ 5 \lambda  }^{\lambda  }-\partial
_{\lambda  }\Gamma _{ 5  5 }^{\lambda  }+\Gamma _{\sigma   5 }^{\lambda  }\Gamma _{ 5 \lambda 
}^{\sigma  }-\Gamma _{\sigma  \lambda  }^{\lambda  }\Gamma _{ 5  5 }^{\sigma  }
\label{ricci-3}
\end{eqnarray}
with new 5-terms from $g_{5\mu}$ and the $\tau$-dependence of
$g_{\mu\nu}$.

\section{Newtonian approximation}
\label{newton}

At low energy the velocity can be approximated as 
\begin{equation}
\dot{x} = \left( c\frac{dt}{d\tau },\frac{d\mathbf{x}}{d\tau },c_{5}\right)
\simeq \left( c\frac{dt}{d\tau }\left( 1,\mathbf{0}\right)
,c_{5}\right)
\end{equation}
so the equations of motion reduce to
\begin{equation}
0 \simeq \ddot{x}^{\mu }+c^{2}\Gamma _{00}^{\mu
} \dot t^{2}+2c c_{5}\Gamma _{50}^{\mu }\dot{t}
+c_{5}^{2}\Gamma _{55}^{\mu } \ .
\end{equation}
Taking $\partial
_{0}g_{\mu \nu } =0$ and $g_{0k} = 0$ in the weak field approximation
\begin{equation}
g_{\mu \nu } =  \eta _{\mu \nu }+h_{\mu \nu }
\end{equation}
we have the new Christoffel symbols
\begin{equation}
\Gamma _{50}^{\mu } =\frac{1}{2c_{5}}g^{\mu 0}\partial _{\tau }h_{00}
\qquad 
\Gamma _{55}^{\mu } =-\frac{1}{2}g^{\mu \nu }\partial _{\nu }h_{55} \ .
\end{equation}
The equations of motion 
\begin{equation}
\ddot{x}^{\mu }  = \frac{1}{2}c^{2}g^{\mu \nu }\partial _{\nu }h_{00}
\dot t^{2}-c\left( g^{\mu 0}\partial _{\tau
}h_{00}\right) \dot t +\frac{1}{2}c_{5}^{2}g^{\mu \nu }\partial
_{\nu }h_{55}
\end{equation}
split into
\begin{equation}
\ddot t = \left( \partial _{\tau }h_{00}\right) \dot t
\qquad \qquad 
\mathbf{\ddot{x}} = \frac{1}{2}c^{2}\dot t
^{2}\nabla h_{00}+\frac{1}{2}c_{5}^{2}\nabla h_{55}
\end{equation}
and writing
\begin{equation}
\frac{d^{2}\mathbf{x}}{d\tau ^{2}}=\frac{d}{d\tau } \left( \frac{dt}{d\tau
}\frac{d\mathbf{x}}{dt}\right)
= \dot t^{2}\frac{d^{2}\mathbf{x}}{dt^{2}}+\left(
\partial _{\tau }h_{00}\right) \dot t\frac{d\mathbf{x}}{dt}
\label{angmomkill}
\end{equation}
we obtain
\begin{equation}
\frac{d^{2}\mathbf{x}}{dt^{2}}  = \frac{1}{2}c^{2}\nabla h_{00}+ \frac{1}{\dot
t^2}\left[ \frac{1}{2}c_{5}^{2}\nabla h_{55}-\left( \partial _{\tau
}h_{00}\right) \dot t\frac{d\mathbf{x}}{dt}\right]  \ . 
\label{x-eqn}
\end{equation}
The unperturbed motion ($\partial _{\tau}h_{\mu \nu }=0$ and $\nabla h_{55}=0$)
recovers Newtonian gravitation
\begin{equation}
\frac{d^{2}\mathbf{x}}{dt^{2}}=-\frac{Gm}{r^{2}}\mathbf{\hat r}=-\nabla \phi \mbox{\qquad}
\qquad 
\phi =-\frac{Gm}{r}
\end{equation}
in the usual manner, with
\begin{equation}
\label{N-appr}
g_{00}=-1-\frac{2}{c^{2}}\phi 
=-\left( 1-\frac{2GM}{ rc^{2}}\right) \ .
\end{equation}
As a perturbation at the Newtonian level, we treat the mass \hbox{$M = M(\tau)$}
as $\tau$-dependent.  As was shown in \cite{antenna}, under certain
circumstances, an accelerated charged particle in SHP electrodynamics may
radiate in such a way that its energy and momentum vary independently, leading
to a change in the particle's mass.  This mass is transferred to the
electromagnetic field, producing mass radiation terms in the mass-energy-momentum
tensor.  Now the $t$ equation
\begin{equation}
\ddot t = \left( \partial _{\tau }h_{00}\right) \dot t = \frac{2G\dot{M}}{rc^{2}} \dot t
\end{equation}
can be solved as
\begin{equation}
\dot t 
= \exp \left( \frac{2G \ \Delta M(\tau) }{rc^{2}}\right)
\end{equation}
where $\Delta M(\tau) = M(\tau) - M(\tau_0)$ is the change in mass from its initial value.
Neglecting $\left( \Delta M \right) ^2$ for small changes in mass and putting $h_{55}=0$, 
the equations for Newtonian gravity 
in plane polar coordinates (with $\dot \theta = d\theta / dt$) become  
\begin{eqnarray}
\frac{d^2r}{dt^2}-r\dot{\theta}^{2} +\frac{2G}{rc^{2}}\frac{dM}{dt}\frac{dr}{dt}
& = & -\frac{GM}{r^{2}} \\
\frac{1}{r}\frac{d}{dt}\left( r^{2}\dot{\theta}\right) +\frac{2G}{c^{2}}\frac{dM}{dt}
\dot{\theta} & = & 0
\end{eqnarray}
after separating the radial and angular parts.  We see that the nonrelativistic 
angular momentum $Mr^2\dot\theta$ is not generally conserved in
this limit, because the transition to $t$ as evolution parameter in
(\ref{angmomkill}) introduced a non-radial velocity-dependent force component.
Neglecting deviation from radial descent (taking $\dot \theta^2 \approx 0$), the radial
equation reduces to 
\begin{equation}
\frac{d^2r}{dt^2} +\frac{2G}{c^{2}}\frac{dM}{dt} \frac{d}{dt} \ln r 
= -\frac{GM}{r^{2}}
\end{equation}
describing an object accelerating in a Newtonian gravitational field with an
additional dissipative term.

\section{Extended Schwarzschild solution}
\label{schw}

We extend the spherically symmetric metric
by permitting $g_{00}=B$ and $ g_{11}=A$ (in the notion of Weinberg
\cite{Weinberg}) to be $\tau$-dependent (but still $t$-independent) and
introducing a fifth metric component $g_{55} = Q$.  Using the spherical symmetry
to put $\theta = \pi / 2$, the Lagrangian becomes  
\begin{equation}
L=\frac{1}{2}\left[ -c^{2}B\left( r,\tau \right) \dot{t}^{2}+A\left( r,\tau
\right) \dot r^{2}+r^{2} \dot{\phi}
^{2}+c_{5}^{2}Q\left( r,\tau \right) \right]
\label{L-S}
\end{equation}
from which we find new non-zero Christoffel symbols
\begin{equation}
\Gamma _{50}^{0}=\dfrac{1}{2c_5}\dfrac{\partial _{\tau }B}{B} 
\qquad 
\Gamma _{55}^{1}  = -\dfrac{1}{2}\dfrac{\partial _{r}Q}{A}\mbox{\qquad}
\Gamma _{15}^{1}=\dfrac{1}{2c_5}\dfrac{\partial _{\tau }A}{A} 
\label{nz-chr}
\end{equation}
and new non-zero Ricci tensor components 
\begin{equation}
\begin{array}{l}
R_{0 5 } =-\dfrac{1}{rc_{ 5 }}\dfrac{\partial _{\tau }B}{B}
\qquad \qquad 
R_{1 5 } =\dfrac{1}{2c_{ 5 }}\left[
\dfrac{\partial _{r}\partial _{\tau }B}{B}+2\dfrac{1}{r}\dfrac{\partial _{\tau
}B}{B}
\right] \strt{18} \\
R_{ 5  5 } 
=\dfrac{1}{2} \left( \dfrac{\partial _{r}^{2}Q}{B}-
\dfrac{\left( \partial _{r}Q\right) \left( \partial _{r}B\right) }{B^{2}}\right)
+\dfrac{1}{2c_{ 5 }^{2}}\left( \dfrac{
\partial _{\tau }B}{B}\right) ^{2}+\dfrac{\partial _{r}Q}{rB}
\end{array}
\end{equation}
suggesting a highly dynamical system.
Taking $M = M(\tau)$ to generalize the Schwarzschild
solution 
\begin{equation}
B(r,\tau) = A^{-1}(r,\tau) = 1-\frac{2GM(\tau)}{rc^{2}}
\end{equation}
we still have $R_{\mu\nu} = \left( R_{\mu\nu}\right)^{4D} = 0$, but 
$R_{05}, R_{15}$, and $R_{55}$ will contain non-vanishing terms
proportional to $dM / d\tau$. 
Setting the terms in $R_{55}$ containing $Q$ to zero we find
\begin{equation}
Q  = \int_{r}^{\infty }dr \ \frac{B}{r^{2}}= \frac{1}{r}\left( 1-
\frac{M\left( \tau \right) G}{c^{2}r}\right)=\frac{1}{2r}\left( 3-B\right)
\end{equation}
leading to the Einstein tensor (\ref{Einstein})
with components
\begin{equation}
\begin{array}{l}
G_{\mu \nu } 
= - \gamma_r^2
\left( 1-\dfrac{MG}{c^{2}r}\right)^{-1} g_{\mu \nu }
\qquad 
G_{55} = \dfrac{1}{r^2}\gamma_r^2
\\
G_{50} = \dfrac{2\gamma_r}{r}
\left( 1- \dfrac{MG}{c^{2}r}\right) ^{-1}
\quad 
G_{51} =- \dfrac{\gamma_r}{r} 
\left( 1- \dfrac{MG}{c^{2}r}\right) ^{-1}
\end{array}
\end{equation}
where $\gamma_r = \dfrac{G}{c_{5}^{2}c^{2}}
\dfrac{dM}{d\tau}\left( 1-\dfrac{2MG}{c^{2}r}\right)^{-1}$.

The equations of motion found from Lagrangian (\ref{L-S}) admit solutions
\begin{equation}
\dot{t}=\frac{1}{B} = \left( 1-\frac{2GM(\tau)}{rc^{2}}\right)^{-1}
\qquad \dot{\phi}=\frac{J}{ r^{2}}
\end{equation}
for constant $J$ and the radial equation
\begin{equation}
\frac{d}{d\tau }\frac{1}{2}\left( -c^{2}\frac{1}{B}+ A\dot{r}^{2} +\frac{J^{2}}{r^{2}}
-c_{5}^{2}Q\right) = -\frac{1}{2}\dot{r}^{2}\partial _{\tau }A - \frac{1}{2}c_{5}^{2}
\partial _{\tau }Q \ .
\end{equation}
We recognize the LHS as the Hamiltonian 
\begin{equation}
K = \frac{1}{2}g_{\mu \nu }\dot{x}^{\mu }\dot{x}^{\nu } - \frac{1}{2}
c_{5}^{2}g_{55}
\end{equation}
in this parameterization, which as we saw in (\ref{non-cons}) is not conserved
when the metric is $\tau$-dependent.  
In equilibrium, when $dM / d\tau = 0$, we recover
\begin{equation}
R_{\alpha\beta} = 0 \qquad  G_{\alpha\beta} = 0 \qquad  j_\alpha =
0\qquad  \frac{dK}{d\tau}  = 0
\end{equation}
describing a massless system (away from the origin), and the radial equation 
admits a first integral solution.

\section{Discussion}

Interactions in SHP electrodynamics form an integrable system in which event
evolution generates an instantaneous current defined over spacetime at $\tau$,
and in turn, these currents induce $\tau$-dependent fields that act on other
events at $\tau$.  We expect that in a similar way, a fully-developed SHP
formulation of general relativity will describe how the instantaneous
distribution of mass and energy at $\tau$ expressed through $T_{\alpha\beta} (x,
\tau)$ induces the local metric $g_{\alpha\beta} (x, \tau)$, which in turn,
determines geodesic equations of motion for any particular event at $x^\mu
(\tau)$.   

Through the simplified example of a $\tau$-dependent mass point, we found that the effect
of the mass shift on the metric leads, by way of the Einstein equations, to mass and energy
radiation into the radial and time directions of spacetime.  We also saw that
this metric induces non-conservation of mass in a particle in extended geodesic
motion.  In this sense, as in SHP electrodynamics, matter may exchange mass
through the medium of local spacetime.
We therefore expect
that SHP general relativity will be a useful computational tool in relativistic
dynamics.

\section*{Acknowledgments}

The author wishes to thank Lawrence P. Horwitz for useful comments on this work,
and to acknowledge his guiding wisdom over many decades.  The author also wishes to
thank Aurora Perez Martinez for providing the impetus for this work.

\bibliography{Land_STARS2019}

\begin{thebibliography}{}

\bibitem [\protect \citeauthoryear {%
Fock%
}{%
Fock%
}{%
{\protect \APACyear {1937}}%
}]{%
Fock}
\APACinsertmetastar {%
Fock}%
\begin{APACrefauthors}%
Fock, V.%
\end{APACrefauthors}%
\unskip\
\newblock
\APACrefYearMonthDay{1937}{}{},
\newblock
\unskip
\newblock
\APACjournalVolNumPages{Phys. Z. Sowjetunion}{12}{}{404--425}.
\newblock
\begin{APACrefURL}
  \url{http://www.neo-classical-physics.info/uploads/3/4/3/6/34363841/fock_-_wkb_and_dirac.pdf}
  \end{APACrefURL}
\PrintBackRefs{\CurrentBib}

\bibitem [\protect \citeauthoryear {%
Gershon%
\ \BBA {} Horwitz%
}{%
Gershon%
\ \BBA {} Horwitz%
}{%
{\protect \APACyear {2009}}%
}]{%
gershon}
\APACinsertmetastar {%
gershon}%
\begin{APACrefauthors}%
Gershon, A.%
\BCBT {}\ \BBA {} Horwitz, L.%
\end{APACrefauthors}%
\unskip\
\newblock
\APACrefYearMonthDay{2009}{}{},
\newblock
\unskip
\newblock
\APACjournalVolNumPages{Journal of Mathematical Physics}{50}{10}{102704}.
\newblock
\begin{APACrefURL} \url{https://doi.org/10.1063/1.3155853} \end{APACrefURL}
\newblock
\begin{APACrefDOI} \doi{10.1063/1.3155853} \end{APACrefDOI}
\PrintBackRefs{\CurrentBib}

\bibitem [\protect \citeauthoryear {%
Horwitz%
}{%
Horwitz%
}{%
{\protect \APACyear {2015}}%
}]{%
rel-qm}
\APACinsertmetastar {%
rel-qm}%
\begin{APACrefauthors}%
Horwitz, L\BPBI P.%
\end{APACrefauthors}%
\unskip\
\newblock
\APACrefYear{2015},
\newblock
\APACrefbtitle {Relativistic Quantum Mechanics} {Relativistic Quantum
  Mechanics}.
\newblock
\APACaddressPublisher{Dordrecht, Netherlands}{Springer}.
\newblock
\begin{APACrefURL} \url{https://www.springer.com/gp/book/9789401772600}
  \end{APACrefURL}
\newblock
\begin{APACrefDOI} \doi{10.1007/978-94-017-7261-7} \end{APACrefDOI}
\PrintBackRefs{\CurrentBib}

\bibitem [\protect \citeauthoryear {%
Horwitz%
}{%
Horwitz%
}{%
{\protect \APACyear {2019}}%
}]{%
SHPGR-1}
\APACinsertmetastar {%
SHPGR-1}%
\begin{APACrefauthors}%
Horwitz, L\BPBI P.%
\end{APACrefauthors}%
\unskip\
\newblock
\APACrefYearMonthDay{2019}{May}{},
\newblock
\unskip
\newblock
\APACjournalVolNumPages{Journal of Physics: Conference Series}{1239}{}{012014}.
\newblock
\begin{APACrefURL}
  \url{https://doi.org/10.1088%2F1742-6596%2F1239%2F1%2F012014}
  \end{APACrefURL}
\newblock
\begin{APACrefDOI} \doi{10.1088/1742-6596/1239/1/012014} \end{APACrefDOI}
\PrintBackRefs{\CurrentBib}

\bibitem [\protect \citeauthoryear {%
Horwitz%
\ \BBA {} Arshansky%
}{%
Horwitz%
\ \BBA {} Arshansky%
}{%
{\protect \APACyear {2018}}%
}]{%
MB}
\APACinsertmetastar {%
MB}%
\begin{APACrefauthors}%
Horwitz, L\BPBI P.%
\BCBT {}\ \BBA {} Arshansky, R\BPBI I.%
\end{APACrefauthors}%
\unskip\
\newblock
\APACrefYear{2018},
\newblock
\APACrefbtitle {Relativistic Many-Body Theory and Statistical Mechanics}
  {Relativistic Many-Body Theory and Statistical Mechanics}.
\newblock
\APACaddressPublisher{}{Morgan \& Claypool Publishers}.
\newblock
\begin{APACrefURL} \url{http://dx.doi.org/10.1088/978-1-6817-4948-8}
  \end{APACrefURL}
\newblock
\begin{APACrefDOI} \doi{10.1088/978-1-6817-4948-8} \end{APACrefDOI}
\PrintBackRefs{\CurrentBib}

\bibitem [\protect \citeauthoryear {%
Horwitz%
\ \BBA {} Piron%
}{%
Horwitz%
\ \BBA {} Piron%
}{%
{\protect \APACyear {1973}}%
}]{%
HP}
\APACinsertmetastar {%
HP}%
\begin{APACrefauthors}%
Horwitz, L\BPBI P.%
\BCBT {}\ \BBA {} Piron, C.%
\end{APACrefauthors}%
\unskip\
\newblock
\APACrefYearMonthDay{1973}{}{},
\newblock
\unskip
\newblock
\APACjournalVolNumPages{Helv. Phys. Acta}{48}{}{316-326}.
\PrintBackRefs{\CurrentBib}

\bibitem [\protect \citeauthoryear {%
Land%
}{%
Land%
}{%
{\protect \APACyear {2019}}%
}]{%
antenna}
\APACinsertmetastar {%
antenna}%
\begin{APACrefauthors}%
Land, M.%
\end{APACrefauthors}%
\unskip\
\newblock
\APACrefYearMonthDay{2019}{May}{},
\newblock
\unskip
\newblock
\APACjournalVolNumPages{Journal of Physics: Conference Series}{1239}{}{012005}.
\newblock
\begin{APACrefURL}
  \url{https://doi.org/10.1088%2F1742-6596%2F1239%2F1%2F012005}
  \end{APACrefURL}
\newblock
\begin{APACrefDOI} \doi{10.1088/1742-6596/1239/1/012005} \end{APACrefDOI}
\PrintBackRefs{\CurrentBib}

\bibitem [\protect \citeauthoryear {%
Land%
\ \BBA {} Horwitz%
}{%
Land%
\ \BBA {} Horwitz%
}{%
{\protect \APACyear {1991}}%
}]{%
lorentz}
\APACinsertmetastar {%
lorentz}%
\begin{APACrefauthors}%
Land, M.%
\BCBT {}\ \BBA {} Horwitz, L\BPBI P.%
\end{APACrefauthors}%
\unskip\
\newblock
\APACrefYearMonthDay{1991}{}{},
\newblock
\unskip
\newblock
\APACjournalVolNumPages{Found. Phys. Lett.}{4}{}{61}.
\PrintBackRefs{\CurrentBib}

\bibitem [\protect \citeauthoryear {%
Saad%
, Horwitz%
\BCBL {}\ \BBA {} Arshansky%
}{%
Saad%
\ \protect \BOthers {.}}{%
{\protect \APACyear {1989}}%
}]{%
saad}
\APACinsertmetastar {%
saad}%
\begin{APACrefauthors}%
Saad, D.%
, Horwitz, L\BPBI P.%
\BCBL {}\ \BBA {} Arshansky, R.%
\end{APACrefauthors}%
\unskip\
\newblock
\APACrefYearMonthDay{1989}{}{},
\newblock
\unskip
\newblock
\APACjournalVolNumPages{Found. Phys.}{19}{}{1125-1149}.
\PrintBackRefs{\CurrentBib}

\bibitem [\protect \citeauthoryear {%
Stueckelberg%
}{%
Stueckelberg%
}{%
{\protect \APACyear {1941}}%
{\protect \APACexlab {{\protect \BCnt {1}}}}}]{%
Stueckelberg-1}
\APACinsertmetastar {%
Stueckelberg-1}%
\begin{APACrefauthors}%
Stueckelberg, E\BPBI C\BPBI G.%
\end{APACrefauthors}%
\unskip\
\newblock
\APACrefYearMonthDay{1941{\protect \BCnt {1}}}{}{},
\newblock
\unskip
\newblock
\APACjournalVolNumPages{Helv. Phys. Acta}{14}{}{321-322}.
\newblock
\APACrefnote{In French}
\PrintBackRefs{\CurrentBib}

\bibitem [\protect \citeauthoryear {%
Stueckelberg%
}{%
Stueckelberg%
}{%
{\protect \APACyear {1941}}%
{\protect \APACexlab {{\protect \BCnt {2}}}}}]{%
Stueckelberg-2}
\APACinsertmetastar {%
Stueckelberg-2}%
\begin{APACrefauthors}%
Stueckelberg, E\BPBI C\BPBI G.%
\end{APACrefauthors}%
\unskip\
\newblock
\APACrefYearMonthDay{1941{\protect \BCnt {2}}}{}{},
\newblock
\unskip
\newblock
\APACjournalVolNumPages{Helv. Phys. Acta}{14}{}{588-594}.
\newblock
\APACrefnote{In French}
\PrintBackRefs{\CurrentBib}

\bibitem [\protect \citeauthoryear {%
Wald%
}{%
Wald%
}{%
{\protect \APACyear {1984}}%
}]{%
Wald}
\APACinsertmetastar {%
Wald}%
\begin{APACrefauthors}%
Wald, R\BPBI M.%
\end{APACrefauthors}%
\unskip\
\newblock
\APACrefYear{1984},
\newblock
\APACrefbtitle {{General relativity}} {{General relativity}}.
\newblock
\APACaddressPublisher{Chicago, IL}{Chicago Univ. Press}.
\newblock
\begin{APACrefURL} \url{https://cds.cern.ch/record/106274} \end{APACrefURL}
\PrintBackRefs{\CurrentBib}

\bibitem [\protect \citeauthoryear {%
Weinberg%
}{%
Weinberg%
}{%
{\protect \APACyear {1972}}%
}]{%
Weinberg}
\APACinsertmetastar {%
Weinberg}%
\begin{APACrefauthors}%
Weinberg, S.%
\end{APACrefauthors}%
\unskip\
\newblock
\APACrefYear{1972},
\newblock
\APACrefbtitle {{Gravitation and Cosmology: Principles and Applications of the
  General Theory of Relativity}} {{Gravitation and Cosmology: Principles and
  Applications of the General Theory of Relativity}}.
\newblock
\APACaddressPublisher{New York, NY}{Wiley}.
\newblock
\begin{APACrefURL} \url{https://cds.cern.ch/record/100595} \end{APACrefURL}
\PrintBackRefs{\CurrentBib}

\end{thebibliography}

\end{document}